\documentclass{elsart5p}
\usepackage{graphicx,natbib}
\usepackage{mathptmx}      
\journal{Social Networks}

\begin{document}
\bibliographystyle{elsart-harv}

\begin{frontmatter}


\title{Affinity Paths and Information Diffusion in Social Networks}
\author{Jos\'e Luis Iribarren\corauthref{cor}}
\address{Instituto de Ingenier\'ia del Conocimiento, Universidad Aut\'{o}noma de Madrid, 28049 Madrid, Spain
}
\ead{jose.iribarren@iic.uam.es}
\author{Esteban Moro\thanksref{moro}}
\address{Instituto de Ciencias Matem\'aticas CSIC-UAM-UC3M-UCM, \\
      Departamento de Matem\'aticas \& GISC, Universidad Carlos III de
      Madrid, 28911, Legan\'es (Madrid), \\
Instituto de Ingenier\'ia del Conocimiento, Universidad Aut\'{o}noma
de Madrid, 28049 Madrid, Spain}
\ead{emoro@math.uc3m.es}
\corauth[cor]{Corresponding author}
\thanks[moro]{E.M. acknowledges partial support from MEC (Spain) through grants
FIS2004-01001, MOSAICO, and a Ram\'on y Cajal contract and Comunidad
de Madrid through grants UC3M-FI-05-077 and SIMUMAT-CM.}

\date{October 25, 2010}

\begin{abstract}{Widespread interest in the diffusion of information through social networks has produced a large number of Social Dynamics models. A majority of them use theoretical hypothesis to explain their diffusion mechanisms while the few empirically based ones average out their measures over many messages of different content. Our empirical research tracking the step-by-step email propagation of an invariable viral marketing message delves into the content impact and has discovered new and striking features. The topology and dynamics of the propagation cascades display patterns not inherited from the email networks carrying the message. Their disconnected, low transitivity, tree-like cascades present positive correlation between their nodes probability to forward the message and the average number of neighbors they target and show increased participants' involvement as the propagation paths length grows. Such patterns not described before, nor replicated by any of the existing models of information diffusion, can be explained if participants make their pass-along decisions based uniquely on local knowledge of their network neighbors affinity with the message content. We prove the plausibility of such mechanism through a stylized, agent-based model that replicates the \emph{Affinity Paths} observed in real information diffusion cascades.}
\end{abstract}

\end{frontmatter}

\section{Introduction and Background}

The discovery of quantitative laws in the collective properties of large numbers of people, for example the birth and death rates or crime frequencies, was one of the factors pushing the development of statistics and led scientists and philosophers to call for some quantitative understanding on how such precise regularities stem from the apparently erratic behavior of individuals. Hobbes, Laplace, Comte, Stuart Mill and many others shared, to a different extent, this line of thought \citep{Ball2004}. The question to investigate was how the interactions between social agents create order in their behavior from an initially disordered state. The basic premise was that agents' repeated interactions should make people more similar since the information exchanges involved led to higher degrees of homogeneity in values, thoughts or preferences. The dynamic nature of the information diffusion, the poor understanding of the human behavior causes and the fact that the agents interactions take place in the thick of complex social networks, made the Social Dynamics problem largely untractable for a long time.

\smallskip

The appearance of new social phenomena related to the Internet (Social Media, Collaborative Filtering, Social Tagging...) whose interactions can be captured in large databases and the tendency of social scientists to move toward the formulation of simplified models and their quantitative analysis, have ushered in an era of scientific research in the field of Social Dynamics \citep{Lazer2009}. Several key questions have been posed: What favors the homogenization process? What hinders it? What are the fundamental interaction mechanisms fostering the adoption of innovations, the spreading of rumors, the evolution towards a dominant opinion or the emergence of trends and fashions?

\smallskip

Initially, the difficulty in obtaining micro-level data on the diffusion of information between individuals, the absence of suitable mathematical algorithms to rigorously analyze the phenomena and the calculation complexity involved in simulations with large real networks limited theoretical advancements to the construction of population average diffusion models based on master or differential equations. Those models were in general borrowed from mathematical epidemiology \citep{Hethcote2000} since it was assumed that information would propagate just like diseases do. However information diffusion research has deeply evolved since step-by-step tracking of interactions through electronic media made detailed diffusion data plentiful (although not necessarily accessible or easy to gather).

\smallskip

The development of the science of complex systems and advancements in the computerized treatment of Social Network Analysis methods have spurred the emergence of a ``new'' science of networks \citep{Watts2004} which provides more robust tools for the scientific treatment of social dynamics processes. As a result scientists realized that information spreading mechanisms vary with the type of information which spawned a rush to develop the appropriate model for each. According to their algorithmic approach those models can be categorized as population-average or network-based. The population-average models assume fully-mixed or homogeneous substrate networks and describe the agents' social dynamic behavior at the aggregate level through differential or master equations. Examples of those are the seminal ``two-step influence model'' of information diffusion by \citet{Katz1955}, the rumor diffusion model of \citet{Daley1965}, the innovations adoption model of \citet{Bass1969}, its stochastic version by \citet{Niu2002}, the minority spreading opinion formation model of \citet{Galam2002}, the innovation diffusion model with influentials and imitators of \citet{Bulte2007} or the percolation-based product lifecycle model of \citet{Frenken2008}. On the other hand, network-based models include the influence of the underlying social network topology by way of agent-based stochastic algorithms. Some examples of them are the classic innovation adoption ``threshold model'' of \citet{Granovetter1978}, the model of diffusion of technological innovations with upgrading costs of \citet{Guardiola2002}, the fads and fashion formation model of \citet{bettencourt2002}, models on the impact of the structural characteristics of a network on innovations diffusion \citep{Jackson2005,Liu2005}, the stochastic model for opinion formation of \citet{Sznajd-Weron2005} or the network variant of the Daley-Kendall rumor model by \citet{Nekovee2007}.

\smallskip

However, this profusion of theoretical models was mainly justified by plausibility arguments and Social Dynamics models based on empirical data are still scarce. A few examples are the referral networks study of \citet{Vilpponen2006} which found that the structure of electronic communication networks is different from that of the traditional interpersonal communication ones, the chain-letter diffusion research of \citet{Liben-Nowell2008} whose strikingly long and narrow spreading chains were attributed to a new mechanism involving asynchronous response times of the forwarders or the study on information diffusion through blogs of \citet{Gomez-Rodriguez2010} which found a core periphery structure in the blogosphere news diffusion network. Nevertheless, all these studies could only trace the propagation of messages with varying content and are unable to discriminate the propagation of individual content items. As a result, none of them could study the impact of the information content on the diffusion processes. While the lack of insight into the content impact would be expected of past century information diffusion research, its absence in more recent literature can only be explained because propagation data at the individual level, being usually proprietary because of its economic value or usage restrictions, is kept under tight wraps and results very hard to obtain.

\smallskip

Our research addresses such shortcoming. Unlike the works cited that study information propagation through the aggregate effect of propagating messages of varying content, ours tracked the precise paths of a viral marketing campaign fixed and invariable message as it spread through an email social network. The message content remained identical through the propagation. This allowed us to scrutinize the individuals' reactions to a particular message instead of just averaged out behavior over diverse information items. By discriminating all factors impacting the participants' spreading patterns from the message content we were able to detect the effects produced by the latter. We found that the message diffusion cascades evolve through a branching process that presents some characteristic and unique patterns unexamined until now although some literature \citep{Leskovec2007,Watts2007a} has shown an inkling of them. We noticed a steady increase in the spreaders' activity parameters as the message gets deeper in the propagation cascades. This surprising pattern can not be observed in empirical experiments collecting propagation data of varying content messages. It can be explained if the cascades growth stems from a mechanism based on the affinity between the message content and the preferences of those receiving it and not on the receiving node neighbors' status or on the underlying social network structure used in many of the current models. We test and validate that hypothesis through a stylized agent-based propagation model. The rest of the article is organized as follows: First we describe the data obtained from our empirical research on real viral marketing campaigns and the control parameters of their messages propagation. Second, we present our findings on the structure and growth patterns of the information cascades. Third we introduce the message affinity propagation model and compare its predictions with the empirical results. The article ends with our conclusions.

\section{Word-Of-Mouth diffusion research}\label{viral}

We tracked and measured the ``word-of-mouth'' diffusion of
viral marketing campaigns ran in eleven European markets which
offered incentives to current subscribers of an IT company online
newsletter to promote new subscriptions through recommendation
emails to friends and colleagues. The campaigns were entirely web
based: banner ads, emails, search engines and the company homepage
drove participants into the campaign site. In it, participants
accessed a referral form to register themselves and enter
the addresses of those to whom they recommended subscribing
the newsletter. The submission of this form triggered a personalized, but otherwise identical, recommendation message with a link to the campaign registration form. The link customized URL was appended with codes allowing to uniquely trace clicks on it to sender and addressee of the corresponding email. The form checked email addresses for syntax correctness and to prevent self
recommendations. Cookies in the participants' email client prevented sending multiple recommendations to the same address\footnote{However, participants with cookies disabled could send multiple referrals to the same person. Thus 183 referrals (0.76\% of total)
were discarded} and improved the user experience by pre-filling the sender's
profile in subsequent sessions. Additionally, the campaign web
server registered a time stamp for each of the process steps
(subscription, recommendations, referral link clicks) and removed
from records referrals to undeliverable email addresses.

\begin{table}
\begin{center}
\begin{tabular}{p{1.6cm}p{1cm}p{1cm}p{1cm}p{1cm}p{1cm}p{0.9cm}r}
\hline \noalign{\smallskip}
\textbf{Market} & $N$ & $N_s$ & $N_v$ & $N_p$ & $Arcs$ & $Casc.$ & $s_{max}$ \\
\noalign{\smallskip} \hline \noalign{\smallskip}
France & 11,758 & 3,247 & 524 & 7,987 & 8,593 & 3,248 & 139  \\
DE+AT & 7,943   &   1,760   &   567 &   5,616   &   6,239   &   1,750   &   146 \\
Spain & 5,260 & 855 & 505 & 3,900 & 4,454 & 843 & 122 \\
Nordic & 2,509 & 530 & 176 & 1,803 & 2,004 & 524 & 34  \\
UK+NL &   2,111   &   521 &   107 &   1,483   &   1,618   &   518 &   25  \\
Italy & 1,602 & 323 & 108 & 1,171 & 1,324 & 319 & 41\\
\hline\noalign{\smallskip}
\textbf{All markets} & \textbf{31,183} & \textbf{7,225} & \textbf{2,002} & \textbf{21,956} & \textbf{24,207} & \textbf{7,188} & \textbf{146} \\
\noalign{\smallskip}\hline
\end{tabular}
\end{center}
\medskip
\caption{\textbf{Campaigns propagation data set:} Count of Total
Nodes ($N$), Seed Nodes ($N_s$), Viral Nodes ($N_v$), Passive Nodes
($N_p$), Total directed links ($Arcs$), and Total of Independent
Cascades ($Casc.$) measured on the campaigns propagation network.
$s_{max}$ is the largest cascade size by its number of nodes.
Quantities in \textbf{All markets} may not add up to the sum of
their column because network partition removes inter-country links.
The number of Seed Nodes ($N_s$) may not coincide with that of
cascades due to cascades merging with one another during the
propagation or because, sometimes, a Seed Node can not be identified
(for example in the case of recommendation reciprocity between two
nodes). Results in some countries are aggregated in homogeneous
markets for statistical significance. Nordic includes DK, FI, NO and
SE.} \label{tab:1}
\end{table}

\smallskip

The incentive offered to recommenders was the possibility of winning laptop computers in a lottery
to be held at the end of the campaign period. Aside from the obvious goal of increasing
participation, the incentive mission was twofold: Firstly,
discourage indiscriminate referrals to prevent spamming-like
behavior and, secondly, ensure legal cover for the tracking of
sender-receiver data. To accomplish such requirement, participation
in the lottery was limited to the so-called successful referrals
defined as the recommendation emails whose recipients clicked on the
coded URL included on them. Thus, the more referral emails sent to
recipients opening them and visiting the campaign site, the higher
the sender's winning odds. More importantly, both sender and
receiver of any successful referral drawn in the lottery were
entitled to receive the lottery prize. Terms and conditions,
accessible from all web site pages and referral emails, specified that
participation in the lottery implied the sender's and receiver's
approval of the campaign registration of their email
transaction details as this was necessary to ensure that both parties could
receive the prize if their referral email was the winning one.
Subscription to the newsletter was not required to participate in
the prize draw. Campaigns ran in each country local language but
were identical otherwise: Identical message, incentive, eligibility
rules, lottery mechanism, campaign duration, web user interface and
tracking processes. This homogeneity of data ensured that behavioral differences between countries were not caused by the campaigns execution but due to the market specifics. It also validates the analysis of  country aggregated results.

\subsection{Campaigns propagation data set}\label{data}

Spurred by the campaign sponsor web site and exogenous online
advertising, a total of 7,225 individuals initiated message
diffusion cascades which grew through viral pass-along driven by
2,002 secondary spreaders. Thus, the viral offering touched another
21,956 passive nodes who did not forward it further. All in all,
31,183 individuals of whom 9,227 were spreaders, received the viral
message. Thus 77\% of the individuals received the message through
the endogenous viral propagation mechanism. The \emph{Cascades
Network} resulting of the message diffusion constitutes a directed
graph with 7,188 independent cascades whose nodes represent
participants linked by 24,207 directed arcs representing the
recommendation emails. We call \emph{Seed Nodes} ($N_s$) the
individuals who spontaneously initiate recommendation cascades from the campaign site without having received a recommendation message from others and \emph{Viral Nodes} ($N_v$) those who forward a previously received message. Table \ref{tab:1} presents the summary data set of the
campaigns message propagation\footnote{The time
dynamics of the message diffusion is covered on a separate
paper}. Unsuccessful emails,
disconnected nodes, nodes with invalid or undeliverable email
addresses, loops and multiple referrals between same nodes were
discarded. In compliance with the sponsor rigorous policy, all
personal information was codified and masked to guarantee the
participants' privacy protection.

\subsection{Cascades Network structural metrics}

\begin{table}
\begin{center}
\begin{tabular}{p{1.5cm}p{0.9cm}p{1cm}p{1cm}p{1.1cm}p{1.1cm}p{0.9cm}c}
\hline \noalign{\smallskip}
\textbf{Market} & $\overline{k}$ & $\sigma_{k}$ & $\overline{k}_{nn}$ & $C$ & $C_{rand}$ & $\overline{\ell}$ & $g_{max}$ \\
\noalign{\smallskip} \hline \noalign{\smallskip}
France & 1.46 & 1.594 & 3.99  & 0.0000 & 0.00012 & 2.164 & 8 \\
DE+AT   &   1.57    &   2.027   &   5.59    &   0.0049  &   0.00020  &   2.671   &   7   \\
Spain &  1.69 & 2.383 & 7.17 & 0.0054 & 0.00032 & 3.287 & 9 \\
Nordic & 1.60 & 1.575 & 4.07 & 0.0077 & 0.00064 & 2.243 & 5 \\
UK+NL   &   1.53    &   1.364   &   3.43    &   0.0112  &   0.00073 &   2.026   &   5   \\
Italy & 1.65 & 1.918 & 5.22 & 0.0234 & 0.00103 & 2.229 & 6 \\
\hline\noalign{\smallskip}
\textbf{All markets} & \textbf{1.55} & \textbf{1.868} &  \textbf{4.97} & \textbf{0.0048} & \textbf{0.00005} & \textbf{2.671} & \textbf{9} \\
\noalign{\smallskip}\hline
\end{tabular}
\end{center}
\medskip
\caption{\textbf{Cascades Network Structural Metrics:}
$\overline{k}$ (total degree) is the average of in- or out-links of
a node, $\sigma_{k}$ its standard deviation, $\overline{k}_{nn}$ the
mean of the nearest neighbors average total degree, $C$ the
\emph{Clustering coefficient}, $C_{rand}=\overline{k}/N$ the
corresponding value for an equivalent random network,
$\overline{\ell}$ the average shortest path length between reachable
nodes (links considered undirected) and $g_{max}$ the maximum number
of steps in directed propagation paths.} \label{metric}
\end{table}

Here we examine differences and similarities between the \emph{Cascades Network} topology and that of the reported email networks through which they propagate. Table \ref{metric} shows the \emph{Cascades Network} structural parameters measured without considering links direction. The cumulative distribution function (c.d.f) of the undirected network total degree $k$ is a power-law $P(k)\sim k^{-2.8}$ whose significant probability of very connected nodes evidences higher heterogeneity than the exponential degree distributions found in some email networks \citep{Guimera2003,Newman2002}. However, their heterogeneity is less marked than that of the email network studied by \citet{Ebel2002b} whose power-law degree distribution (p.d.f.) of exponent $\gamma_{k} = 1.81$ is fatter tailed. Additionally, email networks present positive correlations between the nodes degree at either end of an edge, a property called degree assortativity and measured, according to \citet{Newman2002b}, by the Pearson correlation coefficient. For example, the degree correlation coefficient in the email network of \citet{Guimera2003} is $\rho_k=+0.188$, indication of a correlated network. The equivalent for the \emph{Cascades Network} $\rho_{k}=-0.001$ shows total uncorrelation. Besides, in networks with skewed node degree distributions and degree correlations, such as the email networks, the average connectivity of the network $\overline{k}$ is typically lower than that of the nearest neighbors of a node $\overline{k}_{nn}$. For example in the \citet{Guimera2003} email network, the ratio $\overline{k}_{nn}/\overline{k}$ is approximately 2. Such phenomenon
is responsible for the first neighbors of a node having in average more contacts than such node or, quoting \citet{Feld1991}, for the fact that ``your friends always have more friends than you do.''
Interestingly, this feature is more marked in the \emph{Cascades Network} whose $\overline{k}_{nn}$ to $\overline{k}$ ratio ranges from 2.24 in UK+NL to 4.24 in Spain.

\smallskip

\begin{figure}
\begin{center}
\includegraphics[width=.15\textwidth,viewport= 206 15 400 420 clip=]{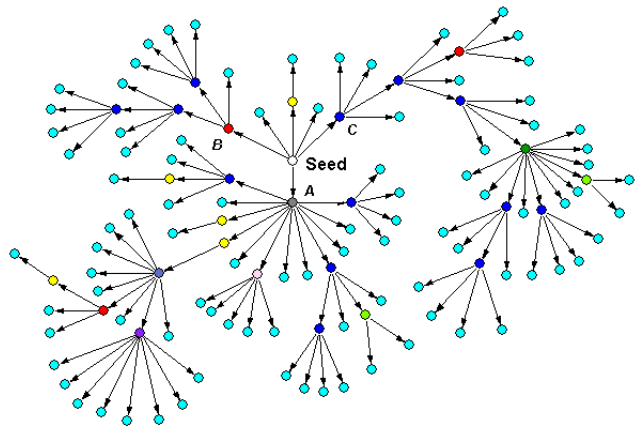}
\end{center}
\caption{\textbf{Tree-like Propagation Cascades:} The viral messages
diffusion graph of our campaigns consists of disconnected cascades
as this one observed in Spain. Its 7 generations and 122 nodes stem
from the node labeled \emph{Seed} and grow through secondary
propagation driven by \emph{Viral Nodes} A, B and C which constitute
$50\%$ of generation 1. Nodes color-coded by their out-degree. The nodes at the end of each path are inactive (out-degree is zero) and do not intervene in the analysis of Section \ref{correlated} which refers to nodes with non-zero in- and out-degree (the \emph{Viral Nodes}).
Notice the tree-like structure devoid of closed paths or triangles
($C=0$). The average total degree of this tree is
$\overline{k}=1.984$ and its largest undirected path (diameter) $d=13$.}\label{cascade}
\end{figure}

Another difference between the \emph{Cascades Network} and the
email networks through which they propagate lies in their transitivity, a
property typical of acquaintance networks whereby two individuals
with a common friend are more likely than average to know each
other. The \emph{Clustering coefficient} $C$, defined as the
fraction of all triangles found in the network relative to the total
number of triads\footnote{A triad is a group of three nodes
connected by two links} measures the transitivity. Table
\ref{metric} shows that our \emph{Cascades Networks} with a
\emph{Clustering coefficient} $C=4.8\times10^{-3}$ for the graph of
\textbf{All markets} are highly intransitive yet ten times more
transitive than an equivalent random network of the same size and
connectivity. In any case, a very low value compared to the range $C
$ [0.15 - 0.60] found in social or email networks \citep{Newman2003c}. Probabilistic considerations show the logic of such feature: since the \emph{Cascades Network} percolates its underlying email network only partially, the dyadic closure that builds clustering in the former must be just a fraction of the one in the latter. As a result our campaigns viral diffusion cascades, like the one in Fig. \ref{cascade}, are almost pure trees.

\smallskip

The last distinctive property of email networks, the \emph{Small World}
or low average shortest path length \citep{Boccaletti2006}, although seemingly present since $\overline\ell=2.67$ (Table \ref{metric}) and lower than that of email networks $\overline{\ell}_{email}\sim3.5$ \citep{Eckmann2004,Guimera2003} is
not comparable with those due to the nature of the \emph{Cascades Network} that, split in many disconnected components, limits paths calculation to reachable pairs of nodes which necessarily yields lower values. The distribution of those cascades size ($s$), like the total degree, is a very skewed power-law whose c.d.f. exponent is $\gamma_s=1.35$. With largest cascade size $s_{max}=146$ nodes, mean size $\overline{s}=4.33$, and $\sigma_s=5.27$, the cascade in Fig. \ref{cascade} is 25 times more likely to appear in our campaigns than in percolation through a random network\footnote{ The tail of the cascade size distribution in large
random networks near the transition to the giant component goes as
$n^c_s\sim s^{-5/2}$ \citep{Albert2002} and the probability of a
cascade of size 122 is $\sim 6.1\times10^{-6}$.}.

\smallskip

In consequence, the viral \emph{Cascades Network} topology lacks all the four key features of email networks (fat tailed node degree distribution, nodes degree correlations, high clustering and the \emph{Small World} property) and can not be formally characterized as a social network. This is quite logical since the viral propagation cascades of diffusion processes far from saturation, such as ours, overlay just sections of the underlying email network and, as a result, can only unveil a small portion of it. Paraphrasing \citet{Liben-Nowell2008} in their study of chain-letters propagation, it is as if ``the progress of the viral messages had a type of stroboscopic effect serving to briefly light up the structure of the global email network.'' Unfortunately, not having any details on the topology of the email network substrate, we can not judge the extent of its influence on the \emph{Cascades Network} topology.

\subsection{Cascades Network Dynamic Parameters}\label{parameters}

While the structure of the undirected cascades is weakly related to that of the email network substrate, the flow of messages in the \emph{Cascades Network} is not (except for the substrate network setting the boundary conditions) and fully depends on the recommendation mechanism.
To study it we will consider the distribution of recommendation emails sent by spreaders of the viral message which offers a better picture of the cascades dynamics than the total node degree $k$ considered so far because 70\% of the network nodes are inactive. This new variable, equivalent to the out-degree of the network nodes, is measured separately for \emph{Seed Nodes} and \emph{Viral Nodes} and designated as $r_s$ and $r_v$ respectively. While most \emph{Viral Nodes} sent just a few recommendations a significant fraction displayed a very intense activity: thus for the ensemble of \textbf{All markets} in our dataset, the mean of the number of recommendations sent by \emph{Viral Nodes}, the so-called \emph{Fanout Coefficient}, was $\overline{r}_v=2.96$ (see Table \ref{dynamical}), its standard
deviation $\sigma_v=7.47$ and the highest number of recommendations sent by a single individual $r_v(max)=72$. Its distribution can be fitted to a fat tailed power-law of the form
\begin{equation}\label{harris}
PL_{\alpha\beta}(r_v)=\frac{H_{\alpha\,\beta}}{\beta+r_v^{\alpha}}
\end{equation}
whose parameters for the \textbf{All markets} network take the
values $H_{\alpha\,\beta}=11.6$, $\alpha=2.83$ and $\beta=10.96$ using Maximum Likelihood Estimation.

\smallskip

\begin{table}
\begin{center}
\begin{tabular}{p{1.5cm}p{0.9cm}p{0.75cm}p{0.75cm}p{1.05cm}p{0.9cm}p{0.75cm}p{0.75cm}c}
\hline\noalign{\smallskip}
\textbf{Market} & $\lambda$ & $\overline{r}_s$ & $\overline{r}_v$ & $\overline{r}_v\textrm{ SEM}$ & $R_0$ & $\overline{s}$ & $\overline{s^*}$ & $\%\,Dev.$ \\
\hline\noalign{\smallskip}
France      & 0.062 & 2.21 & 2.50 & 0.1023 & 0.154 & 3.62 & 3.61 & -0.22 \\
DE+AT   &   0.092   &   2.48    &   3.06    &   0.1155  &   0.281   &   4.54    &   4.45    &   -2.04    \\
Spain       & 0.115 & 3.16 & 3.45 & 0.1909 & 0.400 & 6.24 & 6.23 & -0.20 \\
Nordic      & 0.089 & 2.82 & 2.91 & 0.1836 & 0.259 & 4.79 & 4.81 & +0.31 \\
UK+NL   &   0.067 & 2.49 & 2.87 & 0.2398 & 0.236 & 4.08 & 4.09 & +0.15 \\
Italy       & 0.084 & 2.87 & 2.80 & 0.2301 & 0.236 & 5.02 & 4.76 & -5.20 \\
\hline\noalign{\smallskip}
\textbf{All markets} & \textbf{0.083} & \textbf{2.51} & \textbf{2.96} & \textbf{0.065} & \textbf{0.246} & \textbf{4.34} & \textbf{4.33} & \textbf{-0.30} \\
\noalign{\smallskip}\hline
\end{tabular}
\end{center}
\medskip \caption{\textbf{Cascades growth dynamic parameters:}
\emph{Transmissibility} ($\lambda$), \emph{Fanout Coefficients} of
\emph{Seed} ($\overline{r}_s$) and \emph{Viral} ($\overline{r}_v$)
nodes, Standard error of the \emph{Viral Nodes}
\emph{Fanout coefficient} ($\overline{r}_v\textrm{ SEM}$), Basic Reproductive Number for
secondary spreaders ($R_0$) and average Cascade size ($\overline{s}$) by market as measured in the campaigns. In the last two columns $\overline{s^*}$ is the average Cascade size predicted by the Galton-Watson Branching model Eq. (\ref{avgcascade}) and
$\%\,Dev.$ the deviation of that prediction from
the actual measurements.} \label{dynamical}
\end{table}

We can visualize the cascades of a viral propagation process growing through successive layers, or generations, as nodes reached in one generation resend the message to nodes in the next
generation. The latter nodes constitute the off-spring of the
earlier ones in an evolution of the propagation trees whose
node-level dynamics is well described by the Galton-Watson Branching model\footnote{A markovian model of a population where
each individual in generation $g$ produces in generation $g+1$ a
random number of individuals extracted from the same probability
distribution.} \citep{Harris2002}. Two parameters fully
describe this growth process at the population level: the
aforementioned \emph{Fanout Coefficient} $\overline{r}_v$ and the
message \emph{Transmissibility} $\lambda$ defined as the fraction
of the touched nodes that become secondary spreaders. The
\emph{Transmissibility} results from data in Table \ref{tab:1} as

\begin{equation}
\lambda=\frac{N_v}{N-N_s}
\end{equation}
and both parameters combine to yield the Basic Reproductive Number $R_0$
or average number of secondary recommendations produced by reached
nodes as

\begin{equation}
R_0=\lambda\overline{r}_v
\end{equation}

This number is widely used in mathematical epidemiology \citep{Hethcote2000} to determine the moment when a disease outbreak becomes a self-sustaining epidemic. Thus, if $R_0\geq 1$ the spreading process reaches the \emph{Tipping-point}\footnote{Defined by analogy to phase
transitions in Physics as the process inflection point where
propagation speed accelerates drastically and becomes unstopped so
that the message propagation reaches a very large fraction of the
audience.} an elusive goal that none of our campaigns attained.
Table \ref{dynamical} presents the propagation dynamic
parameters and cascades average size $\overline{s}$ of our campaigns
and their predicted value $\overline{s*}$ for the infinite
propagation limit given by the Galton-Watson Branching model as

\begin{equation}\label{avgcascade}
\overline{s*}=1+\frac{\overline{r}_s}{1-R_0},\qquad R_0\leq1
\end{equation} where $\overline{r}_s$
is the average number of messages sent by \emph{Seed Nodes} and
$R_0$ the viral propagation Basic Reproductive Number. The last
column in Table \ref{dynamical} shows the remarkable accuracy of
the cascades average size predicted by the Galton-Watson Branching
model versus the empirical values.

\section{Patterns of the information cascades growth}\label{patterns}

Despite the Galton-Watson model statistically accurate description of the distribution of cascades at a global level, a detailed study of the \emph{Cascades Network} growth, reveals patterns indicating that viral messages spreading dynamics is quite peculiar. Firstly, we present a node level analysis showing the correlation in the spreading activity of a node with that of its active offspring down the message propagation tree. Secondly, we conduct a generation level analysis on the probability
of the nodes becoming active as a function of their ordinal position in the message diffusion path which shows that viral messages
diffusion propensity increases with distance from the \emph{Seed Node}. Both findings lead to a striking prediction corroborated by the measurements on our viral campaigns: The viral messages diffusion dynamic parameters at the population level are correlated, a fact that has not been observed in other social dynamics processes such as innovations adoption, rumors spreading or opinions propagation. Note that both findings are incompatible with the assumptions in the Galton-Watson model in which the branching mechanism is homogeneous both at the social network level and within the cascades.

\subsection{Correlated spreading of active nodes}\label{correlated}

The first distinctive pattern of the viral messages \emph{Cascades Network} growth is the marked positive correlation of the spreading activity between \emph{Viral Nodes} and their active off-spring. In undirected networks, the nodes total degree correlation is given by the conditional probability $P(k\mid{k'})$ of a node of degree $k$ pointing to a node of degree $k'$. This function is very noisy in finite networks and is usually replaced by the average degree of the nearest neighbors of $k$-degree nodes $\overline{k}_{nn}(k)=\sum_k' k' P(k\mid{k'})$ \citep{Boccaletti2006}. When $\overline{k}_{nn}(k)$ is an increasing function of the degree $k$ the nodes tend to connect to others of similar connectivity and such network, called assortative, displays positive node total-degree correlations.

\begin{figure}
\centering
\includegraphics[width=.58\textwidth,viewport= 18 18 346 213,clip=]{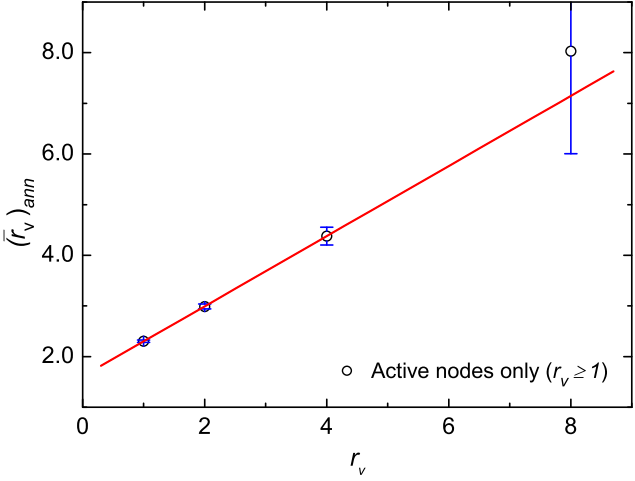}
\caption{\textbf{Active nodes correlated spreading:} Active nearest
neighbors average out-degree $(\overline{r}_v)_{ann}$ (circles) of active nodes $(r_v\geq1)$ in the campaigns as a function of their parents activity $r_v$ in a base two logarithmic binning. The linear fit positive slope (0.69) shows correlation between the spreading activity of a node and that of its active offspring in the propagation tree: the more active a node is, the more active its nearest neighbors in average are.}
\label{assortativity}
\end{figure}

\smallskip

However the active nodes network is directed and instead one should study its out-degree correlation defined as the tendency of nodes to connect with others that have similar out-degrees to themselves. Its formal metric is the \emph{out-assortativity coefficient}\footnote{A convoluted combination of the probability distributions of a link going \emph{out of} a node of out-degree $r_v$, of a link going \emph{into} a node of out-degree $r'_v$ and the joint probability of links to go from a node of out-degree $r_v$ to another of out-degree $r'_v$ \citep{Piraveenan2009}} but considering throughout only the active nodes throughout a simplified analysis of the average out-degree of the \emph{active nearest neighbors} $(\overline{r}_v)_{ann}$ of nodes of out-degree $r_v\geq1$ presented in Fig. \ref{assortativity} suffices to prove that, in terms of the number of recommendations sent in our campaigns, the more active a node is the more prolific in average its progeny is. We studied the out-degree spreading pattern of active nodes in our campaigns (\emph{Seed Nodes} excluded) and found that the activity of a node ($r_v$) correlates with that of its active nearest neighbors. Such correlation implies that the average number of recommendations sent by the active nearest neighbors of a node $(\overline{r}_v)_{ann}$ grows with the number of recommendations $r_v$ that it has sent. The slope of the linear regression of $(\overline{r}_v)_{ann}(r_v)$ is +0.69 indicating strong out-degree correlation. The actual values of $(\overline{r}_v)_{ann}$ range between 1 and 31.33, the mean of their distribution is 2.48 and its standard deviation 2.08.

\smallskip

This very peculiar feature of viral messages diffusion has not been observed on any other type of propagation processes in social networks. We can hypothesize two different explanations of it. One, the increased spreading activity of the active children of a node is a reflection of the out-degree correlation present in the substrate email network. Lacking any data on such network for our campaigns this hypothesis is impossible to verify. Besides, the out-degree positive correlation in the substrate email network merely means that its nodes tend to link to others of similar out-degree but does not by any means indicate that the number of recommendations made by active participants, hence the interest in participating in the campaign, should be a growing function of the number of recommendations made by their parent in the cascade. The other possible explanation, which we adopt, is that the intrinsic mechanism whereby participants in viral marketing campaigns forward the messages involves the sender selecting targets among those of her contacts perceived to be the most receptive to the content of the message being passed-along. The iteration of these target filtering decisions through several generations of senders would lead, in a process akin to targeted search, to focusing the message on groups of individuals genuinely interested on it. Those, in turn, would also be in average more active than their ancestors. The fact that this mechanism has not been observed in other types of information diffusion, such as referral networks \citep{Vilpponen2006}, e-commerce recommendations \citep{Leskovec2007} or email chain-letters \citep{Liben-Nowell2008} may indicate either that the phenomenon is specific of viral marketing messages or that those authors analysis did not isolate the content factor.

\subsection{Diffusion acceleration with path length}

\begin{table}
\centering \scalebox{1.1}{
\begin{tabular}{p{0.5cm}p{1.0cm}p{1.0cm}p{1.0cm}p{1.0cm}p{1.0cm}p{1.0cm}p{0.8cm}}
\hline\textbf{\emph{g}} & $N_g$ & $P_g$ & $(N_v)_g$ & $\lambda_g$ & $(\overline{r}_v)_g$ & $R_g$ & $SEM$  \\
\hline
1 & 18,032 & 0.7527 & 1,398 & 0.0775 & 2.891 & 0.224 & 0.0056 \\
2 &	4,042 & 0.1687 & 393 &	0.0972 & 3.239 & 0.315 & 0.0120 \\
3 &	1,273 &	 0.0531 & 139 & 0.1092 & 2.784 & 0.304 & 0.0228	  \\
4 &	387	& 0.0162 &	40 & 0.1034 & 3.150	& 0.326 & 0.0621 \\
5 &	126	& 0.0053 &	20 & 0.1587 & 3.550	& 0.564 & 0.1804 \\
6 &	71	& 0.0030 &	8 & 0.1127 	& 2.125	& 0.239 & 0.0612 \\
7 &	17	& 0.0007 &	3 & 0.1765 & 2.000 & 0.353 & 0.1765 \\
8 &	6	& 0.0003 &	1 & 0.1667 & 4.000 & 0.667 & 0.0 \\
9 &	4	& 0.0002 &	0 & 0.0	& 0.0 & 0.0	& $N/A$ \\
\hline
\end{tabular}}
\medskip\medskip \caption{\textbf{Distribution of nodes by generation:} Distribution of the nodes touched by the viral message
diffusion in the graph of \textbf{All markets} by ordinal number
$g$ of the position in their diffusion path (generation). $N_g$ is
the number of nodes in generation $g$ and $P_g$ the probability of a node belonging to generation $g\geq1$. $(N_v)_g$
is the number of \emph{Viral Nodes} by generation,  $\lambda_g$ the
probability of nodes in generation $g$ becoming spreaders and
$(\overline{r}_v)_g$ the average number of recommendations sent by
nodes in generation $g$ and $R_g$ the Reproductive Number by generation with $SEM$ its standard error.}\label{generations}
\end{table}

\begin{figure}
\centering
\includegraphics[width=0.43\textwidth,viewport= 20 20 280 220 clip=]{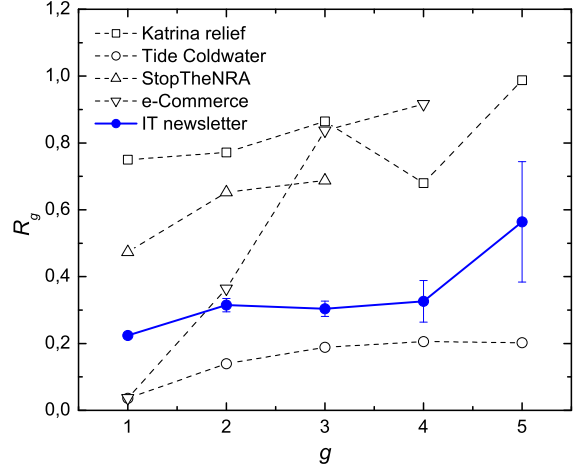}
\caption{\textbf{Diffusion acceleration with path length:}
Reproductive Number by generation $R_g$ in viral messages
propagation. Solid circles with error bars correspond to our IT newsletter campaign. Other data sets (no error bars available): Oxygen Network advocacy portal
collecting contributions for hurricane Katrina relief (squares);
Tide Coldwater campaign for an energy-efficient washing detergent
(empty circles); StopTheNRA, an appeal for gun control launched by
the father of a Columbine shootings victim (upward triangles) per
\citet{Watts2007a}; referrals in e-commerce (downward triangles) per
\citet{Leskovec2007}.
}\label{utility}
\end{figure}

The second characteristic of viral spreading dynamics appears when
measuring the probability of the nodes becoming active spreaders as
a function of their position in the propagation tree. Thus, the
\emph{Transmissibility} by generation $\lambda_g$ in our campaigns
grows in correlation with the ordinal $g$ representing the
individuals' location in the message propagation path. As shown in
Table \ref{generations} for the \textbf{All markets} data,
$\lambda_g$ increases steadily with the generation
($\rho(g|\lambda_g)=0.908$) with parallel growth of the Reproductive Number by generation

\begin{equation}
R_g=\lambda_g(\overline{r}_v)_g=\frac{N_{g+1}}{N_g}
\end{equation}where $N_g$ is the total number of individuals reached at generation $g$. Besides, there is a growth trend for
$(\overline{r}_v)_g$, the \emph{Fanout} by generation which is
visible in our campaigns (Table \ref{generations}) whose
\emph{Fanout} ratio through generations
$(\overline{r}_v)_{g+1}/(\overline{r}_v)_g$ positively correlates
with the generation number ($\rho=0.4$). Those properties of
messages diffusion were detected, but not studied, by \citet{Watts2007a} or \citet{Leskovec2007} as shown in Fig. \ref{utility} along with our campaigns measurements.
As before, we posit that such pattern is due to ``preferential forwarding,'' defined as the spreaders' propensity of passing a message preferentially to neighbors they presume to have more interest, or affinity, for it. Such mechanism results in an increase of the recipients propensity to pass the message along. As a consequence, the message follows network paths such that the \emph{Transmissibility} by generation $\lambda_g$ increases as the propagation progresses. We denominate \emph{Affinity Paths} to the chains of individuals with similar or increasing affinity for the message. They imply some knowledge by message spreaders of their immediate neighbors interests, a local awareness with global impact that leads
to a different class of propagation than that of other Social Dynamics processes. Its consciously driven spreading mechanism causes messages to progress through paths presenting the homophily\footnote{The tendency of individuals to associate and bond with similar others.} properties typical of social networks \citep{Mcpherson2001}. This phenomenon has been observed in the web where, according to \citet {Singla2008} ``there is correlation between preferences and behavior of an individual and those of others in its immediate
circle''.

\subsection{Dynamic Parameters correlation}

\begin{figure}
\centering
\includegraphics*[width=0.45\textwidth,viewport= 18 18 280 210 clip=]{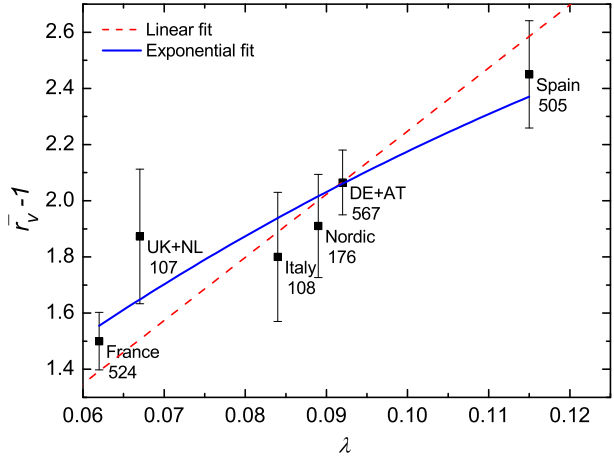}
\caption{\textbf{Dynamic Parameters Correlation:} Correlation
between the population level dynamic parameters
\emph{Transmissibility} ($\lambda$) and \emph{Fanout Coefficient}
($\overline{r}_v$) of our viral campaigns in different markets.
Dotted line is the linear fit to $\overline{r}_v=1+a\lambda$ with
$a=22.48$ and $R^2=0.843$. Solid line is the exponential fit to
$\overline{r}_v=1+b(1-e^{-c\lambda})$ with $b=3.82$, $c=8.44$ and
$R^2=0.818$. Markets position towards the rightmost side of this
``response line'' indicates a higher affinity of the audience with
the campaign message. Number of spreading (active) nodes shown under
market name.}\label{2nd_law}
\end{figure}

As a result of the previous two properties the parameters $\lambda$ and $\overline{r}_v$ are correlated. Let us consider the relationship between the \emph{Fanout Coefficient} and the generation parameters in Table \ref{generations}

\begin{equation}
\overline{r}_v=\frac{\sum_{g=2}N_g}{\sum_{g=1}\lambda_gN_g}=\frac{1-P_g(1)}{\sum_{g=1}\lambda_gP_g}
\end{equation}where $P_g(1)=N_1/\sum_{g=1}N_g=N_s\overline{r}_s/(N-N_s)$ is the probability of an individual to have received the message from a \emph{Seed
Node}. Since $\sum_{g=1}\lambda_gP_g=\lambda$ one obtains the important expression $\lambda\overline{r}_v=1-P_g(1)$ which means that for $\lambda$ and $\overline{r}_v$ to increase simultaneously one must reduce the probability $P_g(1)$ of finding nodes in the first generation or, equivalently, grow longer cascades. Thus, a growing $\lambda_g$ yields longer paths and causes a parallel growth of $\overline{r}_v$. Our campaigns show that the average shortest
path length $(\overline{\ell})$ of the diffusion cascades and the
dynamic parameters are strongly correlated:
$\rho(\overline{\ell}|\overline{r}_v)=0.88$ and
$\rho(\overline{\ell}|\lambda)=0.89$. An increase of the
\emph{Transmissibility} $\lambda$ grows the paths length and the
average number of recommendations made
$\overline{r}_v$ as well. Plotting the dynamic parameters for various markets (Fig. \ref{2nd_law}) their correlation was found to be very strong with a Pearson coefficient $\rho(\lambda|\overline{r}_v)=0.92$. The values of $\lambda$ and $\overline{r}_v$ by country from
Table \ref{dynamical} fit to the
decreasing exponential\footnote{Y intercept set to 1 since
$\overline{r}_v\rightarrow 1$ as $\lambda\rightarrow 0$ because fit
is on active nodes.}

\begin{equation}\label{model}
\overline{r}_v  =  1+b(1-e^{-c\lambda})
\end{equation}which for $\lambda\ll1$, and through a MacLaurin series
expansion of $e^{-c\lambda}$, turns into $\overline{r}_v=1+a\lambda$
($a=bc$). One can consider the slope $a$ of this ``response line'' as the message ``fitness'' with respect to each market. The exponential decrease for large $\lambda$ in Eq. (\ref{model}) is due to the substrate network nodes clustering which limits propagation through saturation and finite size effects.

\smallskip

In principle this correlation between Fanout Coefficient and Transmissibility should invalidate the Galton-Watson model used in Section \ref{parameters}, because that model assumes that those parameters are uncorrelated. However, this is not the case since most of the participants in the campaign appear at very low generation numbers and thus the phenomena observed here is only a significant correction affecting a small fraction of participants.

\section{The Message Affinity Model (MAM)}\label{hypo}

The correlation between the messages propagation dynamic parameters
$\lambda$ and $\overline{r}_v$ and the independence of the nodes
spreading activity from the substrate email network they run upon
are intriguing properties of the viral marketing diffusion processes.
\citet{Watts2007} built a model proving that information propagation
can happen independently of the underlying social network structure
and concluded that ``large cascades of influence are driven not by
the influentials but by a critical mass of easily influenced
individuals.'' However, their model does not explain the dynamic
parameters correlation nor the increase with the generation of the nodes propensity of becoming spreaders. We posit that both features are due to the fact that the decisions of forwarding a viral message and of the number of neighbors to send it to, typically made in a single act by each
forwarding individual, are correlated and that such correlation emerges as a function only of their affinity with the content of the message being spread.

\smallskip

The agent-based Message Affinity Model
(MAM) incorporates that mechanism by assigning to the substrate network nodes a propensity value representing their affinity with the message being forwarded. Furthermore, the model propagation rules combine a variant of the states transition steps of the SIR epidemic model on networks \citep{Pastor-Satorras2001} with the stochastic evolution of a pseudo-markovian\footnote{The Galton-Watson Branching model used in Section \ref{parameters} explains well the growth of the cascades at the average level but fails to predict the activity correlations that appear in the evolution through generations. This is because the Galton-Watson model stochastic process is markovian while, in reality, one node's activity depends on that of its parent.} Galton-Watson Branching model. At any step, the network nodes are in one of the following three states:
\smallskip

\begin{itemize}
        \item \emph{Susceptible (S)}: Node has not received the message
        \item \emph{Informed (I)}: Node is propagating the message
        \item \emph{Refractory (R)}: Node does not spread the message
        anymore
\end{itemize}

\smallskip

Unlike the SIR model, MAM does not use a global probability for the
nodes states transitions. Instead, they stem from the aggregate
decisions that result from the interplay between the nodes
pass-along propensity and the message ``fitness'' to diffuse. Drawn
from a continuous probability density function $p(a)$, the
\emph{Affinity} $a_n\in[0,1]$ of a node represents its propensity to engage in spreading the message. The message fitness to trigger the node activations is represented by their \emph{Affinity Threshold} $A_T\in[0,1]$, the lowest $a_n$ value for which such message can push the node into the \emph{Informed} state: low threshold messages are capable of activating more nodes and are, as a result, forwarded more often than high threshold ones. The process starts by turning a random fraction of the substrate network nodes into the \emph{Informed} state while leaving all others \emph{Susceptible}. From that point onwards the following rules govern the stochastic propagation:
\smallskip

\begin{enumerate}

\item\label{step4} \emph{Susceptible} nodes touched by the message
become \emph{Informed} if their \emph{Affinity} is higher than the
message threshold ($a_n>A_T$) and \emph{Refractory} otherwise while, if touched, \emph{Informed} or \emph{Refractory} nodes stay unchanged.
\smallskip
\item\label{step2} An \emph{Informed} node $n$ forwards a number
of messages $(r_v)_n=(a_n-A_T)\times r $, with $r$ drawn from a
PL distribution. The neighbors receiving those messages are
\smallskip
    \begin{enumerate}
        \item those with highest $a_n$ with probability $(a_n-A_T)$
        \item chosen randomly with probability $1-(a_n-A_T)$
    \end{enumerate}
\smallskip
\item\label{step5} \emph{Informed} nodes become \emph{Refractory}
immediately after spreading the message and the process ends when no
\emph{Informed} nodes are left

\end{enumerate}
\smallskip

The quantity $a_n-A_T$ embodies the interplay between the participants interests and the message content. The choice in Rule (\ref{step2}) of the neighbors that will receive
the message represents the evaluation \emph{Informed} nodes make,
based on their local knowledge, of their neighbors' affinity. It
implies that local knowledge grows with the \emph{Affinity}: nodes
of high $a_n$ are more likely to choose targets with the highest
propensity to pass the message while those with low $a_n$ will
mostly choose their targets randomly.
$A_T$ may vary by individual but, without loss of
generality, we take it constant including all variations in $p(a)$.

\subsection{MAM Simulation Results}\label{validation}

Here we present the result of Monte Carlo simulations of viral messages propagation ran with the MAM model and show that they replicate the patterns observed in real processes. The simulations ran on two substrate networks with the same degree distribution but different structure: the real email network of a Spanish university \citep{Guimera2003} and a synthetic configuration model network built with the Molloy and Reed method \citep{Callaway2001}. They differ in their \emph{Clustering Coefficient} ($C_{email}=0.22$ vs. $C_{conf}=0.014$) and in the fact that the email network node degrees are correlated while the configuration network ones are not. Their nodes \emph{Affinity}, with correlation between nearest neighbors, was drawn from a uniform distribution. The \emph{Cascades Network} resulting from the propagation of messages with \emph{Affinity Threshold} between 0.6 and 0.97 were averaged over 15K cascades with 500 different allocations of the substrate nodes \emph{Affinity}.

\smallskip
\begin{figure}
\centering
\includegraphics[width=.54\textwidth,viewport= 14 18 305 220,clip=]{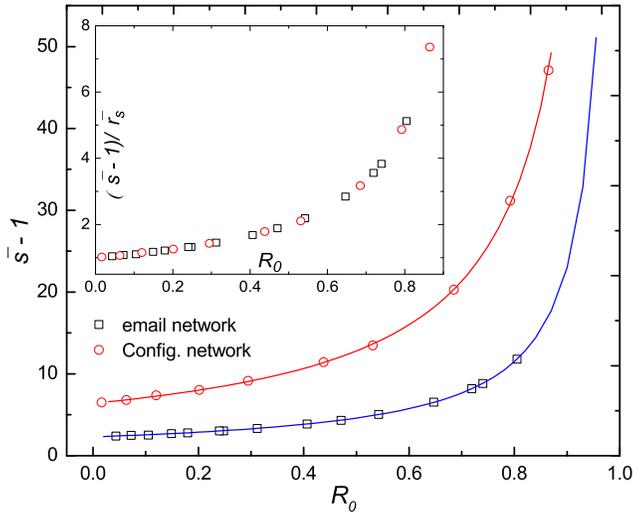}
\caption{\textbf{Cascades Network Average Cascade Size.}
Expected average size of viral cascades for different values of the
Reproductive Number $R_0$ in simulations on the
\textbf{email} (boxes) and \textbf{Config.} (circles) networks with
uniform \emph{Affinity} distribution of mean $\overline{a}_n=0.13$
and $\overline{a}_n=0.17$ respectively. The \emph{Affinity
Threshold} $A_T$ used in the simulations ranges from 0.6 to 0.97 to show the
asymptotic growth for $R_0\sim1$. Solid lines are not a fit but the predictions of Eq.
(\ref{avgcascade}). \textbf{Inset:} Curves collapse when
plotting $(\overline{s}-1)/\overline{r}_s$ against $R_0$ showing the viral
propagation patterns independence of the substrate network
topology.}\label{cascades}
\end{figure}
The simulations generate graphs with a large number of disconnected components that, like those in the real campaigns, feature distributions of Eq. (\ref{harris}) type for both their viral nodes activity $P(r_v)$ and cascades size $P(s)$. The exponents $\gamma_k$ and $\gamma_s$ of their power-laws are in the range 1 - 3 depending on the values of the model parameters nodes \emph{Affinity} ($a_n$) and message \emph{Affinity Threshold} ($A_T$) used. Besides, the average cluster size of the graphs obtained in the simulations follows closely the branching model predictions as shown in Fig. \ref{cascades}. It plots the average size $\overline{s}$ of the propagation network components obtained with different values of the message Affinity Threshold versus their reproductive number $R_0$ for each. The lines are not a fit to the data but the prediction $\overline{s*}$ given by Eq. (\ref{avgcascade}). Notice their remarkable agreement and the fact, shown in the inset, that when the effect of \emph{Seed Nodes} is removed by plotting $(\overline{s}-1)/\overline{r}_s$ the results for the simulations on both substrate networks match exactly. This indicates that as our model predicts, for processes running well below the \emph{Tipping-point} the impact of the substrate network in the cascades average size or the dynamic parameters of the propagation is very low.

\smallskip

\begin{figure}
\centering
\includegraphics[width=.55\textwidth,viewport= 17 19 314 208,clip=]{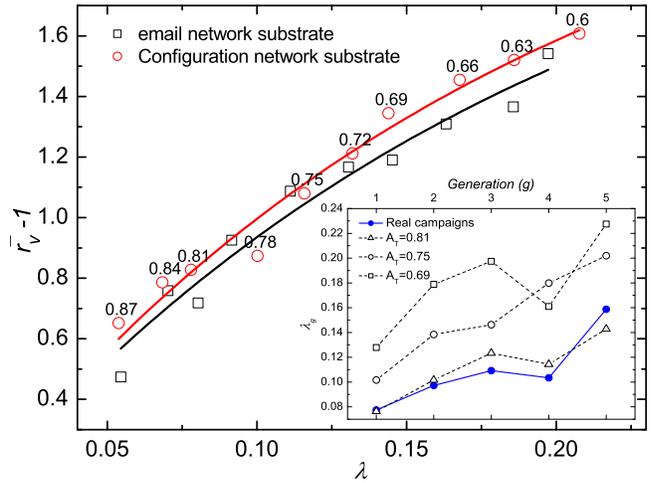}
\caption{\textbf{Correlations in MAM simulations.} Main
panel shows the correlation between dynamic parameters for
simulations on a real email network substrate (boxes) and on an
equivalent configuration model network (circles), both with
uniform distribution of the nodes \emph{Affinity} of mean
$\overline{a}_n = 0.28$. Numbers indicate the message \emph{Affinity
Threshold} ($A_T$) for each simulation. Fits are to Eq. (\ref{model})
with parameters $b_{email}=2.36$, $c_{email}=5.06$ and
$R^2_{email}=0.993$ and $b_{conf}=2.44$, $c_{conf}=5.26$ and $R^2_{conf}=0.976$ for the
respective network substrate. \textbf{Inset:} Evolution of the
\emph{Transmissibility} by generation ($\lambda_g$) for three of the
simulations ($A_T=0.69 - 0.75 - 0.81$) run on the email network
(empty symbols) compared with that of the real
campaigns (full circles). }\label{Affinitymod}
\end{figure}

\smallskip

The plot of the \emph{Cascades
Network} dynamic parameters in the main panel of Fig. \ref{Affinitymod} and their fit to Eq. (\ref{model}) shows how MAM accurately replicates their correlation pattern. This proves that the viral messages propagation patterns are independent of the substrate network structure for low $\lambda$. However their $\overline{r}_v$ values diverge as $\lambda$ grows because the email network clustering and degree correlations accelerate saturation effects and curtail propagation. The diffusion acceleration with path length presented in Fig. \ref{utility} and typical of viral messages propagation is also properly replicated with MAM. The inset of Fig. \ref{model} presents the evolution of $\lambda_g$ with $g$ for simulations on the real email
network (dotted lines) alongside that of our empirical
results. The striking similarity of both up until $g=5$ is quite
significant. The low number of active nodes left in the substrate network beyond that point, renders the statistics of the results unreliable. The same pattern (not shown) appears for simulations on the configuration model network. The growth of $\lambda_g$ can not prevent the propagation process ending. In fact for $R_g<1$, $\lambda_g$ is a
probability below unity applied at each subsequent generation to an ever shrinking cohort of nodes. As proved by the Branching Process theory, the cascades inevitably reach a point where there is no new offspring and they die off. Actually, even for $R_g>1$ the cascades extinction has a non-zero probability that increases with the heterogeneity of the participants' activity distribution \citep{Harris2002}.

\section{Conclusions and Discussion}\label{conclusions}

We tracked and analyzed the structure and growth dynamics of the propagation
network created by the diffusion of a content-controlled message in real viral marketing campaigns driven through email forwarding. The resulting \emph{Cascades Network}, formed by almost pure trees of very low clustering, shows two striking dynamical patterns not observed so far in other Social Dynamics processes like rumor spreading, innovations adoption or email chain-letters. First, there is positive correlation between the spreading nodes activity level as measured by their out-degree and that of their active off-spring and, second, the propensity of nodes reached by the message to becoming spreaders, the \emph{Transmissibility} $\lambda$, grows with those nodes depth in the propagation path. These novel properties can only be detected by scrutinizing the propagation of messages of fixed and identical content. The scarcity of such type of data may explain why they have remained unobserved until now. The discovered patterns have two remarkable consequences. On the one hand, the dynamic parameters \emph{Transmissibility} and \emph{Fanout Coefficient} for a given message across different markets are correlated. On the other, the topology of the email network underlying the propagation has limited influence on the \emph{Cascades Network} although its features are compatible with the structure of the substrate email network that conditions their formation.

\smallskip

Our explanation of all those peculiarities stems from the mechanism driving the messages propagation which involves the affinity of the campaign participants with the content of the message. Participants would make a simultaneous and conscious decision of spreading it or not and to whom which leads to a the positive correlation between the probability of becoming a spreader after receiving the message and the average number of messages forwarded. This decision would result from a single intrinsic property of the nodes in the substrate network, their affinity with the message being passed-along. Besides, the dynamic parameters by generation $\lambda_g$ and $(\overline{r}_g)$ tend to grow with $g$ since the choice of targets to forward the message to is based on the participants' awareness of their neighbors' affinity with it. Such mechanism steers the message through paths of increased affinity termed \emph{Affinity Paths}.

\smallskip

This hypothesis is tested through an agent-based model (MAM) that replicates the patterns discovered and validates the proposed Affinity-driven information diffusion mechanism. It combines a stochastic branching process with propagation rules that create cascades of touched nodes by taking the substrate network nodes message awareness through a sequence of \emph{Susceptible}, \emph{Informed} and \emph{Reluctant} states. The MAM uses just two control parameters: the \emph{Affinity} distribution $p(a)$ of the substrate network nodes to assign them an affinity value between 0 (message is not sent) and 1 (message will certainly be forwarded) and the \emph{Affinity Threshold} $A_T$ representing the message fitness to be passed-around. As the model runs through a substrate network list of edges, the interplay between $A_T$ and the nodes \emph{Affinity} generates cascades with all the expected features while providing a glimpse into the substrate network topology. The empirical analysis and the theoretical model validate our conclusion that the mechanism driving viral marketing messages propagation results from the affinity between the campaign participants' preferences and the messages content. In fact, the viral cascades features depend more on the individuals' reaction to the message than on the substrate network topology. However, we could not verify this conclusion empirically since the structure of our campaigns substrate network being unknown, a comparison between the \emph{Cascades Network} and the substrate email network was impossible. Also, MAM does not replicate the merging of cascades that occurs near the \emph{Tipping-point} as it assumes that \emph{Seed Nodes} are planted in a boundless network and far apart of each other to avoid propagation clashing. Finally, MAM only runs on undirected and fully connected networks.

%

\bibliography{Master}

\begin{thebibliography}{39}
\expandafter\ifx\csname natexlab\endcsname\relax\def\natexlab#1{#1}\fi
\expandafter\ifx\csname url\endcsname\relax
  \def\url#1{\texttt{#1}}\fi
\expandafter\ifx\csname urlprefix\endcsname\relax\def\urlprefix{URL }\fi

\bibitem[{Albert and Barab\'{a}si(2002)}]{Albert2002}
Albert, R., Barab\'{a}si, A.-L., 2002. Statistical mechanics of complex
  networks. Rev. Modern Phys. 74, 47--97.

\bibitem[{Ball(2004)}]{Ball2004}
Ball, P., 2004. Critical Mass: How One Thing Leads to Another. Farrar, Strauss
  and Giroux, London, UK.

\bibitem[{Bass(1969)}]{Bass1969}
Bass, F.~M., 1969. A new product growth model for consumer durables. Management
  Science 15, 121--227.

\bibitem[{Bettencourt(2002)}]{bettencourt2002}
Bettencourt, L. M.~A., 2002. From boom to bust and back again: the complex
  dynamics of trends and fashions. cond-mat/0212267.

\bibitem[{Boccaletti et~al.(2006)Boccaletti, Latora, Moreno, Chavez, and
  Hwang}]{Boccaletti2006}
Boccaletti, S., Latora, V., Moreno, Y., Chavez, M., Hwang, D.-U., 2006. Complex
  networks: Structure and dynamics. Physics Reports 424, 175--308.

\bibitem[{Callaway et~al.(2001)Callaway, Hopcroft, Kleinberg, Newman, and
  Strogatz}]{Callaway2001}
Callaway, D.~S., Hopcroft, J.~E., Kleinberg, J.~M., Newman, M.~E.~J., Strogatz,
  S.~H., 2001. Are randomly grown graphs really random? Phys. Rev. E 64,
  041902.

\bibitem[{Daley and Kendall(1965)}]{Daley1965}
Daley, D.~J., Kendall, D.~G., 1965. Stochastic rumours. IMA Journal of Applied
  Mathematics 1(1), 42--55.

\bibitem[{Ebel et~al.(2002)Ebel, Mielsch, and Bornholdt}]{Ebel2002b}
Ebel, H., Mielsch, L.-I., Bornholdt, S., 2002. Scale-free topology of e-mail
  networks. Phys. Rev. E 66, 035103(R).

\bibitem[{Eckmann et~al.(2004)Eckmann, Moses, and Sergi}]{Eckmann2004}
Eckmann, J.-P., Moses, E., Sergi, D., 2004. Entropy of dialogues creates
  coherent structures in e-mail traffic. Proc. Natl. Acad. Sci. USA 101,
  14333--14337.

\bibitem[{Feld(1991)}]{Feld1991}
Feld, S., 1991. Why your friends have more friends than you do. American
  Journal of Sociology 96, 1464--1447.

\bibitem[{Frenken et~al.(2008)Frenken, Silverberg, and Valente}]{Frenken2008}
Frenken, K., Silverberg, G., Valente, M., 2008. A percolation model of the
  product lifecycle. Tech. rep., United Nations University - UNU-MERIT.

\bibitem[{Galam(2002)}]{Galam2002}
Galam, S., 2002. Modelling rumors: the no plane pentagon french hoax case.
  Physica A 320, 571--580.

\bibitem[{Gomez-Rodriguez et~al.(2010)Gomez-Rodriguez, Leskovec, and
  Krause}]{Gomez-Rodriguez2010}
Gomez-Rodriguez, M., Leskovec, J., Krause, A., 2010. Inferring networks of
  diffusion and influence. In: The 16th ACM SIGKDD Conference on Knowledge
  Discovery and Data Mining (KDD).

\bibitem[{Granovetter(1978)}]{Granovetter1978}
Granovetter, M., 1978. Threshold models of collective behavior. American
  Journal of Sociology 83~(6), 1420--1443.

\bibitem[{Guardiola et~al.(2002)Guardiola, D\'iaz-Guilera, P\'erez, Arenas, and
  Llas}]{Guardiola2002}
Guardiola, X., D\'iaz-Guilera, A., P\'erez, C., Arenas, A., Llas, M., 2002.
  Modelling diffusion of innovations in a social network. Physical Review E 66,
  026121.

\bibitem[{Guimer\'{a} et~al.(2003)Guimer\'{a}, Danon, D\'{i}az-Guilera, Giralt,
  and Arenas}]{Guimera2003}
Guimer\'{a}, R., Danon, L., D\'{i}az-Guilera, A., Giralt, F., Arenas, A., 2003.
  Self-similar community structure in a network of human interactions. Phys.
  Rev. E 68, 065103(R).

\bibitem[{Harris(2002)}]{Harris2002}
Harris, T.~E., 2002. The Theory of Branching Processes. Springer-Verlag,
  Berlin.

\bibitem[{Hethcote(2000)}]{Hethcote2000}
Hethcote, H.~W., 2000. The mathematics of infectious diseases. SIAM Review
  42~(4), 599--653.

\bibitem[{Jackson and Yariv(2005)}]{Jackson2005}
Jackson, M., Yariv, L., 2005. Diffusion on social networks. \'Economie Publique
  16, 3--16.

\bibitem[{Katz and Lazarsfeld(1955)}]{Katz1955}
Katz, E., Lazarsfeld, P.~F., 1955. Personal influence; the part played by
  people in the flow of mass communications. Glencoe, Ill: Free Press.

\bibitem[{Lazer et~al.(2009)Lazer, Pentland, Adamic, Aral, Barab\'asi, Brewer,
  Christakis, Contractor, Fowle, Gutmann, Jebara, King, Macy, Roy, and
  Alstyne}]{Lazer2009}
Lazer, D., Pentland, A., Adamic, L., Aral, S., Barab\'asi, A.-L., Brewer, D.,
  Christakis, N., Contractor, N., Fowle, J., Gutmann, M., Jebara, T., King, G.,
  Macy, M., Roy, D., Alstyne, M.~V., 2009. Computational social science.
  Science 323~(5915), 721--723.

\bibitem[{Leskovec et~al.(2007)Leskovec, Adamic, and Huberman}]{Leskovec2007}
Leskovec, J., Adamic, L., Huberman, B., 2007. The dynamics of viral marketing.
  ACM Transactions on the Web 1, 1.

\bibitem[{Liben-Nowell and Kleinberg(2008)}]{Liben-Nowell2008}
Liben-Nowell, D., Kleinberg, J., 2008. Tracing information flow on a global
  scale using internet chain-letter data. Proc. Natl. Acad. Sci. USA 105~(12),
  4633--4638.

\bibitem[{Liu et~al.(2005)Liu, Madhavan, and Sudharshan}]{Liu2005}
Liu, B. S.-C., Madhavan, R., Sudharshan, D., 2005. Diffunet: The impact of
  network structure on diffusion of innovation. European Journal of Innovation
  Management 8~(2), 240--262.

\bibitem[{McPherson et~al.(2001)McPherson, Smith-Lovin, and
  Cook}]{Mcpherson2001}
McPherson, M., Smith-Lovin, L., Cook, J.~M., 2001. Birds of a feather:
  Homophily in social networks. Annual Review of Sociology 27, 415--444.

\bibitem[{Nekovee et~al.(2007)Nekovee, Moreno, Bianconi, and
  Marsili}]{Nekovee2007}
Nekovee, M., Moreno, Y., Bianconi, G., Marsili, M., 2007. Theory of rumour
  spreading in complex social networks. Physica A 374, 457--470.

\bibitem[{Newman(2002)}]{Newman2002b}
Newman, M. E.~J., 2002. Assortative mixing in networks. Phys. Rev. Lett. 89,
  208701.

\bibitem[{Newman et~al.(2002)Newman, Forrest, and Balthrop}]{Newman2002}
Newman, M. E.~J., Forrest, S., Balthrop, J., 2002. Email networks and the
  spread of computer viruses. Phys. Rev. E 69, 026113.

\bibitem[{Newman and Park(2003)}]{Newman2003c}
Newman, M. E.~J., Park, J., 2003. Why social networks are different from other
  types of networks. Phys. Rev. E 68, 036122.

\bibitem[{Niu(2002)}]{Niu2002}
Niu, S.-C., 2002. A stochastic formulation of the bass model of new-product
  diffusion. Review of Marketing Science Working Papers 1~(4), 1.

\bibitem[{Pastor-Satorras and Vespignani(2001)}]{Pastor-Satorras2001}
Pastor-Satorras, R., Vespignani, A., May 2001. Epidemic dynamics and endemic
  states in complex networks. Phys. Rev. E 63~(6), 066117.

\bibitem[{Piraveenan et~al.(2009)Piraveenan, Prokopenko, and
  Zomaya}]{Piraveenan2009}
Piraveenan, M., Prokopenko, M., Zomaya, A., 2009. Assortative mixing in
  directed biological networks. IEEE Transactions on Computational Biology and
  Bioinformatics.

\bibitem[{Singla and Richardson(2008)}]{Singla2008}
Singla, P., Richardson, M., 2008. Yes, there is a correlation - from social
  networks to personal behavior on the web. In: Proceeding of WWW'2008.

\bibitem[{Sznajd-Weron(2005)}]{Sznajd-Weron2005}
Sznajd-Weron, K., 2005. Sznajd model and its applications. Act. Phys. Pol. B
  36~(8).

\bibitem[{{Van den Bulte} and Joshi(2007)}]{Bulte2007}
{Van den Bulte}, C., Joshi, Y.~V., 2007. New product diffusion with
  influentials and imitators. Marketing Science 26~(3), 400--421.

\bibitem[{Vilpponen et~al.(2006)Vilpponen, Winter, and
  Sundqvist}]{Vilpponen2006}
Vilpponen, A., Winter, S., Sundqvist, S., 2006. Electronic word-of-mouth in
  online environments: exploring referral network structure and adoption
  behavior. Journal of Interactive Advertising 6~(2).

\bibitem[{Watts(2004)}]{Watts2004}
Watts, D.~J., 2004. The "new" science of networks. Annual Review of Sociology
  30, 243--270.

\bibitem[{Watts and Dodds(2007)}]{Watts2007}
Watts, D.~J., Dodds, P.~S., 2007. Influentials, networks and public opinion
  formation. Journal of Consumer Research 34, 441--458.

\bibitem[{Watts and Peretti(2007)}]{Watts2007a}
Watts, D.~J., Peretti, J., 2007. Viral marketing for the real world. Harvard
  Business Review F0705A.

\end{thebibliography}

\end{document}